\documentclass[pra]{revtex4}

\usepackage{graphics,graphicx}
\usepackage{mathrsfs}
\usepackage{subfigure}
\usepackage{color}

\newcommand\beq{\begin{equation}}
\newcommand\enq{\end{equation}}
\def\beqa{\begin{eqnarray}}
\def\eeqa{\end{eqnarray}}
\def\ba{\begin{array}}
\def\ea{\end{array}}

\newcommand{\f}{\frac}
\newcommand{\tf}[2]{\ensuremath{#1/#2}}
\newcommand{\pa}[1]{\ensuremath{\left(#1\right)}}
\newcommand{\paa}[1]{\ensuremath{\left\{#1\right\}}}

\newcommand{\paf}[2]{\ensuremath{\left(\f{#1}{#2}\right)}}

\def\de{\delta}

\def\th{\theta}
\def\Th{\Theta}

\def\Om{\Omega}
\def\om{\omega}

\newcommand{\wt}[1]{\ensuremath{\widetilde{#1}}}

\newcommand{\Int}[2]{\ensuremath{\int\limits_{#1}^{#2}}}

\newcommand{\sul}[2]{\ensuremath{\sum\limits_{#1}^{#2}}}
\newcommand{\ppl}[2]{\ensuremath{\prod\limits_{#1}^{#2}}}

\newcommand{\s}[1]{\ensuremath{\sin\pa{#1}}}
\renewcommand{\c}[1]{\ensuremath{\cos\pa{#1}}}
\newcommand{\shc}[1]{\ensuremath{\mathrm{shc}\pa{#1}} }

\newcommand{\ex}[1]{\ensuremath{\mathrm{e}^{#1}}}

\newcommand{\abs}[1]{\ensuremath{\left| #1 \right|}}

\newcommand{\moy}[1]{\ensuremath{\langle #1 \rangle}}
\newcommand{\moyy}[1]{\ensuremath{[ #1 ]}}
\newcommand{\dd}{\mathrm{d}}
\newcommand{\eee}[1]{\ensuremath{\mathrm{#1}}}

\newcommand{\intn}[2]{\ensuremath{[\![ \, #1 \,;\, #2 \,]\!]}}
\renewcommand{\v}[1]{\ensuremath{\mathbf{#1}}}
\renewcommand{\r}{\ensuremath{\v{r}}}

\long\def\symbolfootnote[#1]#2{\begingroup%
\def\thefootnote{\fnsymbol{footnote}}\footnote[#1]{#2}\endgroup}
\def\@fnsymbol#1{\ifcase#1\or *\or \dagger\or \ddagger\or \mathchar "278\or \mathchar "27B\or \|\or **\or \dagger\dagger \or \ddagger\ddagger \else\@ctrerr\fi\relax}

\begin{document}

\title{On 4-point correlation functions in simple polymer models}

\author{Johannes-Geert Hagmann$^{1,\ddagger}$\symbolfootnote[0]{$^{\ddagger}$The two authors have equally contributed to this work.}, Karol K. Kozlowski$^{1,\ddagger}$, Nikos Theodorakopoulos$^2$ and Michel Peyrard$^1$}
\address{$^1$ Universit\'e de Lyon, Laboratoire de Physique, Ecole Normale Sup\'erieure de Lyon, CNRS, 46 All\'ee d'Italie, 69364 Lyon, France}
\address{$^2$ Theoretical and Physical Chemistry Institute, National Hellenic Research Foundation, Vasileos Constantinou 48, 116 35 Athens, Greece}

\date{\today}

\begin{abstract}
\noindent
We derive an exact formula for the covariance of cartesian distances in two simple polymer models, the freely-jointed chain and a discrete flexible model with nearest-neighbor interaction. We show that even in the interaction-free case correlations exist as long as the two distances at least partially share the same segments.
For the interacting case, we demonstrate that the naive expectation of increasing correlations with increasing interaction strength
only holds in a finite range of values. Some suggestions for future single-molecule experiments are made.
\end{abstract}

\maketitle

\section{Introduction}
With the advent of new powerful tools for imaging and manipulation, the detection of single molecules and the characterization of their dynamics is a rapidly growing and developing field in experimental biophysics. Recently, the possibility to track molecular motion at the level of a single molecule has received substantial attention. Indeed, such methods can in principle provide a glimpse of a biological molecule ''at work'' (see \cite{ha} for a recent review). The complexity and variety of interactions in biomolecules makes it hard to study quantitatively the biomolecular dynamics
through a theoretical model that ought to reproduce most of the experimental observations. It is thus in computer simulations of empirical models \cite{karplus,vangunsteren} that the most considerable quantitative studies of biomolecular dynamics can be found.
The challenge of analyzing simulation results is usually somewhat opposite to the case of single-molecule experiments.
On the one hand, in experiments one attempts to extract as much information on the dynamics as possible from a small number of accessible observables.
On the other hand, for the interpretation of numerical simulations one needs to develop schemes allowing to reduce a large amount of data resulting from the numerous degrees of freedom to a meaningful subset.

\noindent Obviously, if not guided by prior motivation, the ways to achieve this reduction are manifold. A large number of such schemes have been proposed, involving linear and nonlinear concatenation of degrees of freedom (see e.g. \cite{brown} for a comparison of common approaches). Unfortunately, very few of these methods provide testable predictions on observables that are directly accessible in experiments. Hence,  their potential for the interpretation of experiments probing biomolecular motion is limited.

\noindent In this work, we explore an alternative direction for probing biomolecular motion by introducing an experimentally testable and hence confutable framework allowing to characterize cooperative motion in the simulation of empirical models for biomolecules. More precisely, we introduce 4-point correlation functions of distances as a new measure that can serve to compare results from multiple dye single-molecule F\"orster-Resonance Energy Transfer (FRET) with simulations and analytical results on models of biopolymers. While most measures characterizing the dynamics of biomolecules start directly from numerical observations, in this work we build the theory of 4-point correlation functions in the limiting cases of exactly solvable polymer models which are subsequently validated by numerical experiments. As the implementation of the formalism is not limited to the underlying model, applications to more complex models which can only be treated numerically will be reported in future work. The results on the simple models presented here already reveal important issues regarding the influence of interaction strength as well as the geometry that are relevant to the interpretation of future experiments.

\noindent The article is organized as follows. In section \ref{section2} we outline the interest of choosing a 4-point correlation function for characterizing the collective motion of biomolecules as a generalization of the present 2-point FRET-based experiments. Focussing on analytically solvable models, we calculate both analytically and numerically the 4-point correlations functions of two simple polymer models: the freely-jointed chain in section \ref{section3} and a discrete flexible model with nearest-neighbor interaction in section \ref{kratky}. In both cases we study the convergence of covariances as estimated in computational studies or single-molecule experiments from finite-time averages towards the ensemble average determined by our exact calculations. We conclude by a discussion of our findings in respect to future single molecule experiments in section \ref{outlook}.

\section{Design of 4-point correlation functions for biomolecules}
\label{section2}
Present day single-molecule FRET techniques can retrieve the time-resolved distance between a donor and acceptor molecule. The donor/acceptor is a dye molecule attached to the biomolecule or an unit of the biomolecule itself, whose dynamics reveal the motion of the underlying molecule. Recent highlights in experimental single molecule FRET setups include the high resolution long-time scale observation of protein/ligand and protein dynamics \cite{yang1,yang2}, and the observation of fluctuations by three-color FRET in the DNA four-way junction \cite{hw}. The measurement of FRET signals with more than two dye molecules in a single biomolecule has the advantage of yielding more than one distance at a time. Still, such approaches are technically difficult to realize, and require to overcome several difficulties, e.g. the selective dye labelling within the probe molecule and a more demanding theoretical framework to extract distance information from the multiple efficiency signals \cite{watrob} to name but a few.

\noindent Despite these experimental and theoretical challenges, it is constructive to anticipate what information is to be gained from such experiments,
keeping in mind the rapid progress in this field over the past years. We address the question: how do {\em two} distances revealed by FRET measurement
correlate in biomolecules? In other words, given two distinct pairs of points $(A,B)$ and $(C,D)$ in the single molecule, to what extent can correlation be expected when one meaures the time-dependent distances 
$R_1=\abs{\v{R}_{AB}}$ and $R_2=\abs{\v{R}_{CD}}$?

\noindent The covariance of distances constitutes a simple measurement of this correlated motion. For the two time-dependent
observables, its definition reads
 \begin{eqnarray}
\mathcal{C}({R}_1^2,{R}_2^2,T)= \overline{\left({R}^2_1(t) - \overline{{R}^2_1}  \right)} \overline{ \left(R^2_2(t) - \overline{ {R}^2_2 } \right)} =
\overline{R_1^2(t)R_2^2(t)}-\overline{R_1^2}\ \overline{R_2^2} \ \ \ .
\end{eqnarray}
There $\overline{  A(t)}  =\frac{1}{T}\int_{t_0}^{t_0+T}dt\ A(t)$ denotes the time average of the observable $A(t)$ over a time period $T$. The average is assumed to start at the beginning of the measurement when $t=t_0$, consequently, the value of $\mathcal{C}(R_1^2,R_2^2,T)$ in principle also depends on the choice of $t_0$. This dependence only vanishes in the limit of long $T$ or averaging over different $t_0$. $\mathcal{C}({R}^2_1,{R}^2_2,T)$ is positive if the two signals simultaneously increase/decrease and negative if they vary in the opposite sense. Finally, the covariance is zero if the fluctuations of these distances are uncorrelated. We further denote by $C(R_1^2,R_2^2)$ the ensemble average value of the covariance to which $\mathcal{C}(R_1^2,R_2^2,T)$ should converge in the limit $T\rightarrow \infty$. We will moreover consider a reduction of the experimental setting, where the maximum distance for which a FRET signal can be obtained is limited, and calculate the covariance for any distance.\\In the following sections, we calculate the covariance $C(R_1^2,R_2^2)$ analytically and numerically for two simple polymer models, the
freely-jointed chain which is the simplest model, and a discrete flexible model which corresponds to polymer chains with nearest-neighbor interactions. 

\section{Freely-jointed chain}
\label{section3}
\begin{figure}
\centering
\includegraphics[width=2.7in]{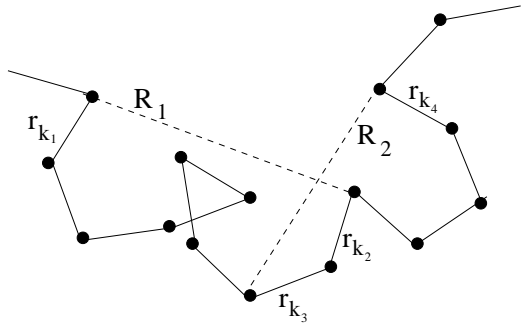}
\caption{scheme of the notations used for the freely jointed chain}
\label{fig1}
\end{figure}

Let us now derive the distance covariance $C(R^2_1,R^2_2)$ from ensemble averages for the freely jointed chain representing a classical polymer model without interactions. We consider a chain of $N$ monomers of fixed length $b$  with no explicit interactions between them. Each monomer vector $\mathbf{r}_i$ ($|\mathbf{r}_i|=b$) is labelled by a discrete index. We will denote the distances to be considered for correlation by
$\mathbf{R}_1=\sum_{i=k_1}^{k_2}\mathbf{r}_i,\ \mathbf{R}_2=\sum_{i=k_3}^{k_4}\mathbf{r}_i
$, where the order $k_1\le k_2$, $k_3\le k_4$, $k_i\in \lbrace 1,2,...,N \rbrace$ is assumed. This notation is illustrated in figure \ref{fig1}. Most of the classical results on this chain can be found in \cite{doi}. The $N$-segment probability distribution function factorizes due to the lack of explicit interaction. It reads
\begin{eqnarray}
P(\lbrace \mathbf{r}_i \rbrace)&=&\Pi_{i=1}^{N} p(\mathbf{r}_i)=\Pi_{i=1}^{N} \frac{\delta(|\mathbf{r}_i|-b)}{4\pi b^2}  \label{fjcpdf} \ \ \ \ .
\end{eqnarray}

\subsection{Analytical results}
%
%
%
%
%
%
The purpose of this calculation is to understand how non-vanishing 4-point correlations can arise between distances in polymers
even if there is no specific interaction that could be responsible for cooperative effects. We study a correlator between the squared distances
\begin{eqnarray}
C(R_1^2,R_2^2)&\equiv&\langle \mathbf{R}_1^2 \mathbf{R}_2^2\rangle-\langle \mathbf{R}_1^2 \rangle \langle\mathbf{R}_2^2\rangle \ \ \ \ .
\end{eqnarray}
\noindent As the probability distribution function factorizes, one has $C\pa{R_1^2,R_2^2}=0$ whenever $\lbrace k_1,...,k_2 \rbrace\cap \lbrace k_3,...,k_4 \rbrace=\emptyset$. Consequently, we only consider the case $\lbrace k_1,...,k_2 \rbrace\cap \lbrace k_2,...,k_4 \rbrace\neq\emptyset$.
 We split the summation in a way allowing to cancel the factorizable terms:
\begin{eqnarray}
\langle \mathbf{R}_1^2 \mathbf{R}_2^2\rangle&=&\langle\left[ \sum_{ij=k_1}^{k_3-1}
(\mathbf{r}_i\cdot\mathbf{r}_j) + \sum_{ij=k_3}^{k_2}
  (\mathbf{r}_i\cdot\mathbf{r}_j)\right]\cdot\left[
\sum_{kl=k_3}^{k_2}
(\mathbf{r}_k\cdot\mathbf{r}_l) + \sum_{kl=k_2+1}^{k_4}
(\mathbf{r}_k\cdot\mathbf{r}_l) \right]\rangle \; .
\end{eqnarray}
There, the non-factorizable contribution arises from the overlapping sums.
Summing up the factorizable terms, we have
\begin{eqnarray}
C(R_1^2,R_2^2)&=& \langle\left( \sum_{ij=k_3,\  i\neq j}^{k_2} (\mathbf{r}_i\cdot\mathbf{r}_j)\right)^2 \rangle  =
\frac{2}{3}b^4(k_2-k_3+1)(k_2-k_3) \label{rescov}\; ,
\end{eqnarray}
after an explicit integration in spherical coordinates. From this simple result, we draw the following conclusions.\\
\textit{i)} A non-zero covariance arises for geometrical overlap in the distances $R_1,R_2$ along the sequence, i.e. if $R_1$ and $R_2$ share at least \underline{two} monomers, the covariance of the squared distances scales as the square of the overlap $\zeta=k_2-k_3+1$. The result is independent of the size of the system, or the overall length of the distances $R_1,R_2$ themselves as it should be. The distances have to share at least two monomers as for as single shared monomer the auto-correlation in $\langle R_1^2 R_2^2\rangle$ is still contained in the offset $\langle R_1^2\rangle\langle R_2^2\rangle$. From a physical point of view, the result is plausible as no information from neighboring segments can be transmitted through the chain owing to the absence of interacting forces. Only the part of the topology which is shared among the distances $R_1,R_2$, i.e. the $\zeta$ overlapping segments, can contribute to a non-vanishing correlation.
\\
\textit{ii)} The value of the covariance is positive semidefinite; there is on average no anti-correlated motion in the chain.

\noindent In the interacting case this picture should change. In particular, the result should involve the whole lengths $R_1,R_2$ at least for moderate 
interactions. Yet, strong interaction will tend to constraint the motion of the polymer into a single configuration. In this case,
$\langle R_1^2 R_2^2\rangle$=$\langle R_1^2\rangle\langle R_2^2\rangle$, and $C(R_1^2,R_2^2)=0$.
\subsection{Numerical results}\label{num1}
\begin{figure}
\centering
\includegraphics[width=2.4in]{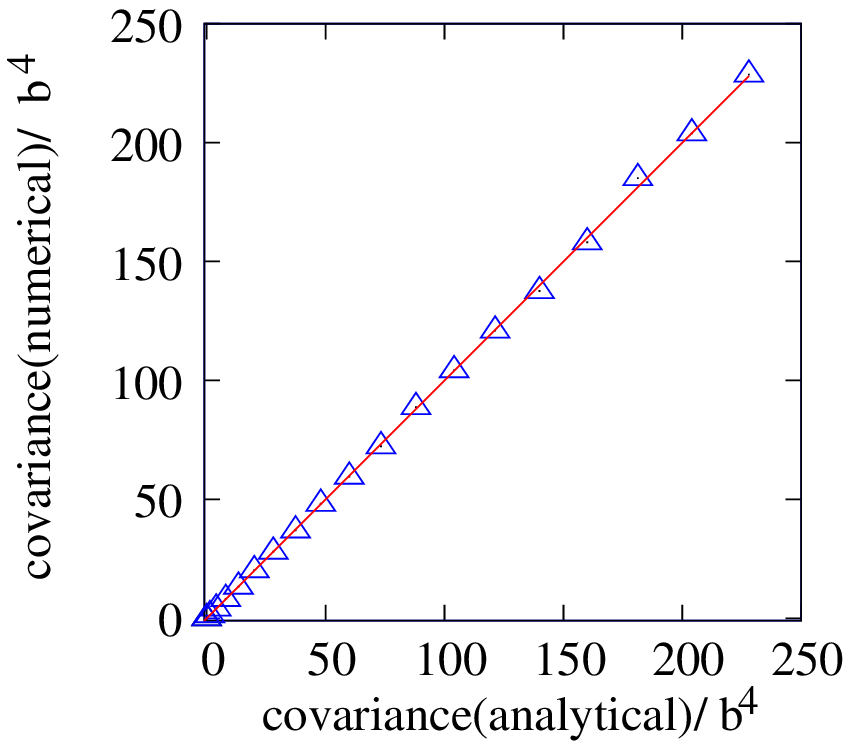}
\includegraphics[width=2.3in]{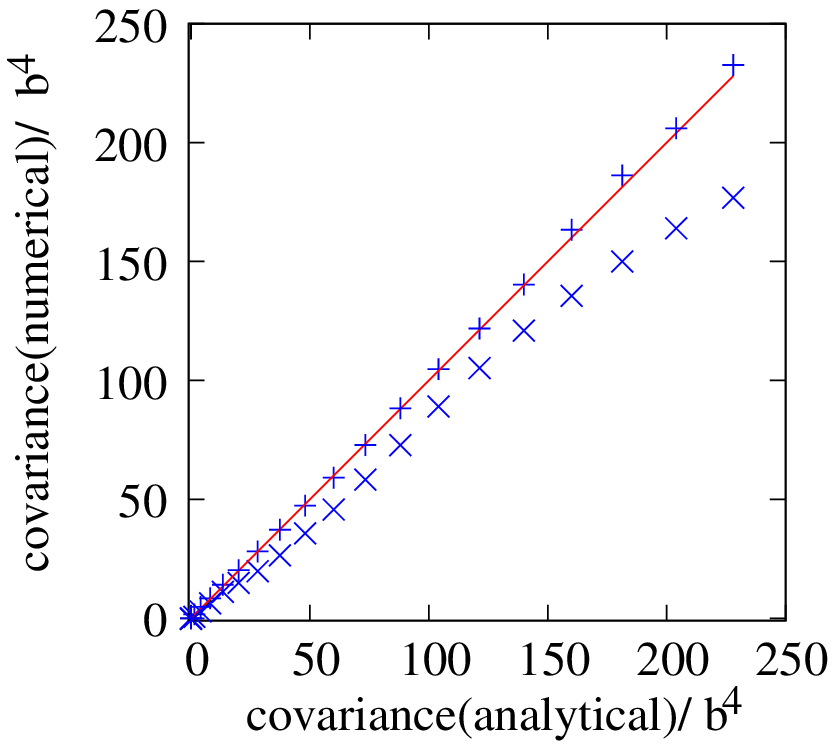}
\caption{\textit{Left:} Covariance calculated according to (\ref{rescov}) and numerical results (simple sampling, $N_c=5\cdot 10^4$, $N=50$, $\zeta=0,1,..19$). The straight line $f(x)=x$ is a guide to the eye. \textit{Right:} Covariance calculated according to (\ref{rescov}) and numerical results (molecular dynamics, see text). $+$: $m=1.0$, $\times$: $m=100.0$, $\zeta=0,1,..19$.}
\label{fig3}
\end{figure}

\paragraph{ensemble averages}
We verified (\ref{rescov}) by simple sampling on an ensemble average of $N_c$ randomly generated freely-jointed chains according to the probability distribution (\ref{fjcpdf}) with a fixed number of total segments $N$ and fixed monomer length $b$. The left-hand side of figure \ref{fig3} shows a comparison of the covariance values $C(R_1^2,R_2^2)$ between the ensemble averages obtained from $N_c=5\cdot10^4$ chains of length $N=50$ segments and the analytical result for varying number of overlap monomers $\zeta=0,1,...,19$. The spacing between the points increases as the covariances scales with $\zeta^2$ in the overlap.
\paragraph{time averages}
Besides a static ensemble average, we can also estimate dynamical time averages. This can be achieved by implementing a chain of mass points with rigid links for which we numerically solve the constrained equations of motion. An adequate representation for the freely-jointed chain can be obtained by simulating a chain of $N$ harmonic springs using constrained molecular dynamics. Here, we use the popular Brooks-Brunger-Karplus algorithm for Langevin
 dynamics \cite{bbk} combined with RATTLE \cite{rattle} to solve the equations of motion of $N+1$ particles joined by harmonic springs, applying the constraints $|\mathbf{r}_i|=b$ at each time-step of the integration. Note that such an approach has
to be taken with care for several reasons.
\\ \textit{i)} Ideally, such a chain would be massless to avoid biasing the trajectory of an individual monomer by a
collective inertial motion. This case can only be approximated if the chain is strongly coupled to a heat bath so that inertial
 effects are overdamped, and the time interval between the configurations recorded to calculate averages is sufficiently
 long so that the monomer orientations can be effectively considered as uncoupled.
\\ \textit{ii)} It has recently been pointed out that applying constraints to a stochastic heat bath leads to temperature dependent equilibrium
bond lengths \cite{franklin}. If the covariance was to be evaluated at different temperatures, as the bond length enters with $b^4$, a correction
needs to be applied to the integration algorithm.
\\All observables of the simulation are expressed in dimensionless units. We simulate $N+1=51$ point masses of equal mass $m=1.0$ joint by harmonic springs (uniform spring constant $D=1.0$) and an equilibrium bond length between two masses $b=1.0$. Using a time step of $\Delta t=0.1$ and a friction constant $\gamma=0.1$, we have $\gamma\Delta t \ll 1$ and $\Delta t \ll (2\pi)/\sqrt{D/m}$. The initial velocities were taken from a Maxwell-Boltzmann distribution, and the chain is
 thermalized at $k_BT=0.05$ for $10^5$ time units in order to guarantee the loss of memory of the initial condition (a partially stretched configuration).  Configurations were recorded every
$100$ time units to compute averages. For a constant temperature trajectory of length $10^6$ time units, the temporally averaged kinetic energy closely approached the canonical expectation value.
The right-hand side of figure \ref{fig3} illustrates the effects of inertia on the comparison with the analytical results. For small masses, the overdamped dynamics of the constrained harmonic chain approximate the phase space sampling of the freely-jointed chain quite well at least on the timescale of the simulation.
However, the results for a larger mass exhibits a strong deviation. The latter is observed to be more pronounced when the overlap $\zeta$ is located at the center of the chain (data not shown) compared to the case of an overlap located at the ends.  Indeed, the free end is only affected by inertial motion from one side.
\subsection{Relation of the distance covariance to the third order susceptibility in spin systems}
\label{relation}
In classical spin systems, fourth-order correlation functions arise naturally when analyzing higher order susceptibilities as response functions to an external field.
To see the analogy with the calculation of the higher-order correlations with polymers, we define the Hamiltonian of the system to be
\begin{eqnarray}
H=H_0+\mathbf{h}_1\cdot\mathbf{O}_1+\mathbf{h}_2\cdot\mathbf{O}_2
\end{eqnarray}
where $H_0$ is an unperturbed Hamiltonian of some spin chain and $\mathbf{O}_1,\mathbf{O}_2$ are two observables of the system coupling to the external fields $\v{h}_1,\v{h}_2$. We now can introduce third-order susceptibility $\chi^{(3,*)}_{ijkl}$ associated with mixed derivatives of the on-site free energy $f$ in the two fields $\v{h}_1,\v{h}_2$:
\begin{eqnarray}
\chi^{(3,*)}_{ijkl}&=&-\frac{\partial^4 f}{\partial h_{1,i}   \partial h_{1,j}\partial h_{2,k}\partial h_{2,l}}\mid_{h_1=h_2=0} \ \ \ \ , \ \ \ \ f=-\lim_{N\rightarrow\infty}\frac{k_BT}{N}\log Z \ .
\end{eqnarray}
In this particular case we get, for Hamiltonians with vanishing first order moments in the observables $\mathbf{O}_1$ and $\mathbf{O}_2$,
\begin{eqnarray}
\chi^{(3,*)}_{iijj}&=&(k_BT)^{-3}\lim_{N\rightarrow\infty}\frac{1}{N}\left(\langle O_{1,i} O_{1,i} O_{2,j} O_{2,j} \rangle -\langle O_{1,i} O_{1,i} \rangle\langle O_{2,j} O_{2,j} \rangle -2 \langle O_{1,i} O_{2,j} \rangle^2\right) \ \ \ .
\end{eqnarray}
$\chi^{(3,*)}_{iijj}$ can be related to $C(R_1^2,R_2^2)$. After a summation over the indices $i,j$, the first two terms within the bracket in the previous expression correspond to $C(R_1^2,R_2^2)$ provided that the identification $O\leftrightarrow R$ is made. The extra term $\langle R_{1,i} R_{2,j}\rangle$ has however
no correspondence in the previous calculation. In contrast to the fourth-order correlation we have defined above, calculating the fourth-order
 susceptibility would require not only the information on the distances $|\mathbf{R}|$, but also on their orientation.
Such an information cannot be accessed by FRET techniques. Hence, in practice, the measurement only allows to infer the covariance which is not an intensive quantity in the overlap.
\\
\noindent We have seen that the analogy with spin systems provides a relation between the covariance measure with a generalized susceptibility in a spin system. For the moment, our calculations deal with the simplest possible case of a polymer chain without interaction. In the following section, we go beyond this first order, calculating the covariance in a polymer model
with nearest-neighbor interactions. This allows us to assess the qualitative and quantitative change of the results upon introducing interactions.
The analogy with spin systems will also prove very useful in this context, as the starting point for our calculation are exact results known for spin
 systems.
\section{Discrete flexible chain}
\label{kratky}
A classical system for studying polymers is the worm-like chain model (WLC). In this section, we study the behavior of a discrete flexible chain with nearest-neighbor interactions \cite{kp} which in the continuum limit and for large coupling constants yields the WLC model. A related model has been analyzed in the context of loop formation in double stranded DNA \cite{marko}. In contrast to the freely-jointed
 chain which represents a non-interacting system of stiff rods, this has the additional physical feature of penalizing local
  bending by a harmonic force deriving from the Hamiltonian
\begin{eqnarray}
H&=&-\epsilon\sum_{i=1}^{N-1}(\mathbf{r}_i\mathbf{r}_{i+1}-b^2)
\end{eqnarray}
where $|\mathbf{r}_i|=b$ and $\epsilon$ is a uniform coupling constant. A rescaling of the Hamiltonian $H\rightarrow H-\epsilon(N-1)b^2$, makes the system
equivalent to the zero-field limit of the classical one-dimensional Heisenberg chain with uniform coupling and open boundaries. The latter has been extensively studied, see e.g. \cite{mattis} for a physical perspective.\\In 1994 Nakamura and Takahashi \cite{takahashi} reported the nonlinear susceptibility for a uniform field in this system. These results can thus serve as a starting point for the calculation of the distance covariance in an interacting polymer model. However, in order to be fully applicable, these should be generalized along the lines described in the following section.
\subsection{Analytical results}
As the calculation of the covariance for the discrete flexible chain is more involved than for the freely-jointed chain, we start by an outline of the
 procedure.
\\ We are interested in the covariances
\begin{eqnarray}
C(R_1^2,R_2^2)&=&\langle \mathbf{R}_1^2 \mathbf{R}_2^2\rangle-\langle \mathbf{R}_1^2 \rangle \langle\mathbf{R}_2^2\rangle \label{fullcov}
\end{eqnarray}
with $\mathbf{R}_1=\sum_{i=k_1}^{k_2} {\bf r_i},\ \mathbf{R}_2=\sum_{i=k_3}^{k_4}  {\bf r_i} $ as defined previously. Averages are taken with respect to the canonical
probability distribution function bound to the rescaled discrete flexible chain Hamiltonian
\begin{eqnarray}
H=-K k_BT\sum_{i=1}^{N-1}\mathbf{\hat{\mathbf{r}}_i}\cdot\mathbf{\hat{\mathbf{r}}_{i+1}}
\end{eqnarray}
with $K=\epsilon/(k_BT b^2)$ and $\mathbf{r}=b\hat{\mathbf{r}}$. The squared distance correlation functions  are deduced by using the isotropy of the zero-field Hamiltonian
and Fisher's result \cite{fisher} of local correlations in the model (see also Appendix \ref{appendix1}). One obtains
\begin{eqnarray}
\langle\mathbf{R}_1^2\rangle&=&3\langle {R}_1^z R_1^z\rangle= \sum_{ij=k_1}^{k_2} u^{-|i-j|}=\frac{2(a+1)-(a+3)u^a}{1-u}-1-\frac{u(1-u^{a-1})}{(1-u)^2} \; ,
\label{fisher}
\end{eqnarray}
where $u=\coth(K)-K^{-1}$ and $a=k_2+1-k_1$. However, the correlator of the two squared distances%
\begin{eqnarray}
\langle \mathbf{R}_1^2 \mathbf{R}_2^2\rangle&=&\sum_{ij=k_1}^{k_2}\sum_{kl=k_3}^{k_4}\langle (\mathbf{r}_i \cdot \mathbf{r}_j)(\mathbf{r}_k \cdot  \mathbf{r}_l) \rangle \label{r12}
\end{eqnarray}
cannot be directly inferred from previous results and deserves special attention. The two sums run over different domains, such that the use of permutation symmetry is subject to restrictions. The isotropy of the interactions can be used to reduce the number of correlators to be evaluated to two, i.e. $\langle {r}^x_i {r}^x_j {r}^z_k {r}^z_l \rangle$ and  $\langle {r}^z_i {r}^z_j {r}^z_k {r}^z_l \rangle$. All the other correlators are deduced from a rotation of the reference frame (the latter being a symmetry of the Hamiltonian). $\langle {r}^z_i {r}^z_j {r}^z_k {r}^z_l \rangle$ has been derived in \cite{takahashi} (see also \cite{tomita} cited therein), whereas  $\langle {r}^x_i {r}^x_j {r}^z_k {r}^z_l \rangle$ is not known as it usually does not arise in problems dealing with magnetism where the applied field is uniform.
Combining these observation, we get
\begin{eqnarray}
\langle \mathbf{R}_1^2 \mathbf{R}_2^2\rangle&=&\sum_{ij=k_1}^{k_2}\sum_{kl=k_3}^{k_4} \left(6\langle {r}^x_i {r}^x_j {r}^z_k {r}^z_l \rangle+3\langle {r}^z_i {r}^z_j {r}^z_k {r}^z_l \rangle\right) \ \ \ \ .
\end{eqnarray}
In \ref{appendix1}, we derive the correlators $\langle r^x_i r^x_j r^z_k r^z_l \rangle$ and  $\langle r^x_i r^z_j r^x_k r^z_l \rangle$ for $i\le j\le k \le l$. The result reads
\begin{eqnarray}
\langle r^x_i r^x_j r^z_k r^z_l \rangle&=&\langle r^z_i r^z_j r^x_k r^x_l \rangle =u^{j-i+l-k}\left(\frac{1}{15}v^{k-j} +w\frac{v^{k-j}-1}{v-1}\right) \ \ \ \ ,\label{m1} \\
\langle r^x_i r^z_j r^x_k r^z_l \rangle&=&\langle r^x_i r^z_j r^z_k r^x_l \rangle=\frac{1}{15}u^{j-i+l-k}t^{k-j} \label{m2} \; .
\end{eqnarray}
We also recall here the result of \cite{takahashi} on $\langle r^z_i r^z_j r^z_k r^z_l \rangle$:
\beq
\langle r^z_i r^z_j r^z_k r^z_l \rangle =\frac{1}{9} u^{j-i} \left(\frac{4}{5}v^{k-j}+1 \right)u^{l-k} \label{m3} \; .
\enq
In the above formulae, we have introduced
\begin{eqnarray}
v&=&1-3\frac{u}{K} \ \ \ \ , \ \ \ \  w=\frac{1}{3} \left( K^{-1}\coth(K) - K^{-2}\right)\ \ \ \ , \nonumber \\
t&=&\frac{1}{K^2}\left(3+K^2-3K \coth(K) \right)\  \ \ \ .
\end{eqnarray}
With these exact results, we can decompose the summations involved in (\ref{fullcov}) by symbolic computation \cite{mathematica}. The results for varying coupling constant $K$ and chain overlap $\zeta$ are presented in the following section along with the results obtained from direct numerical computation.

\subsection{Numerical results}
\begin{figure}
\centering
\includegraphics[width=2.7in]{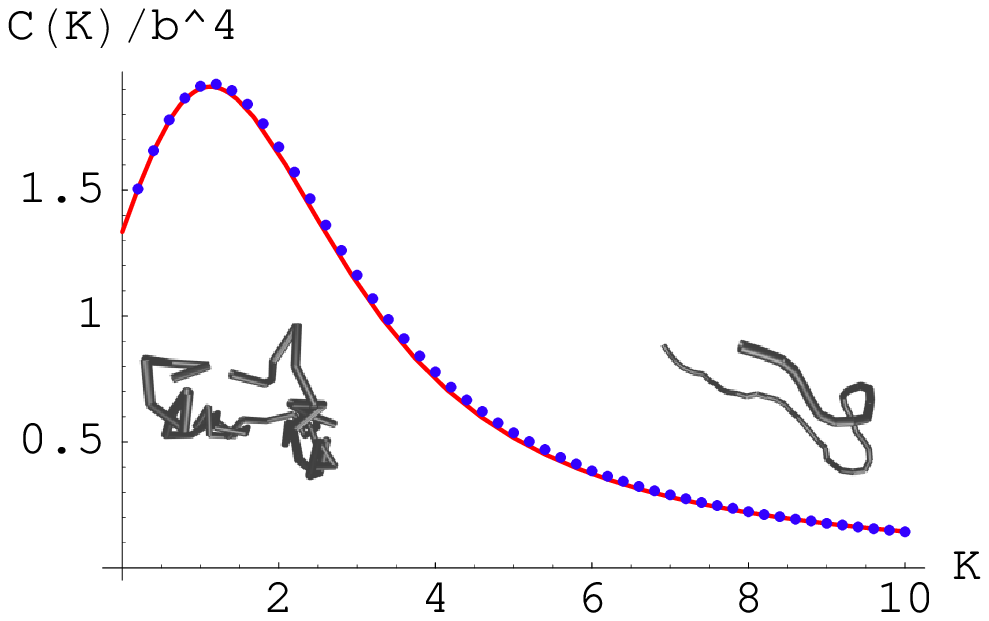}
\includegraphics[width=2.7in]{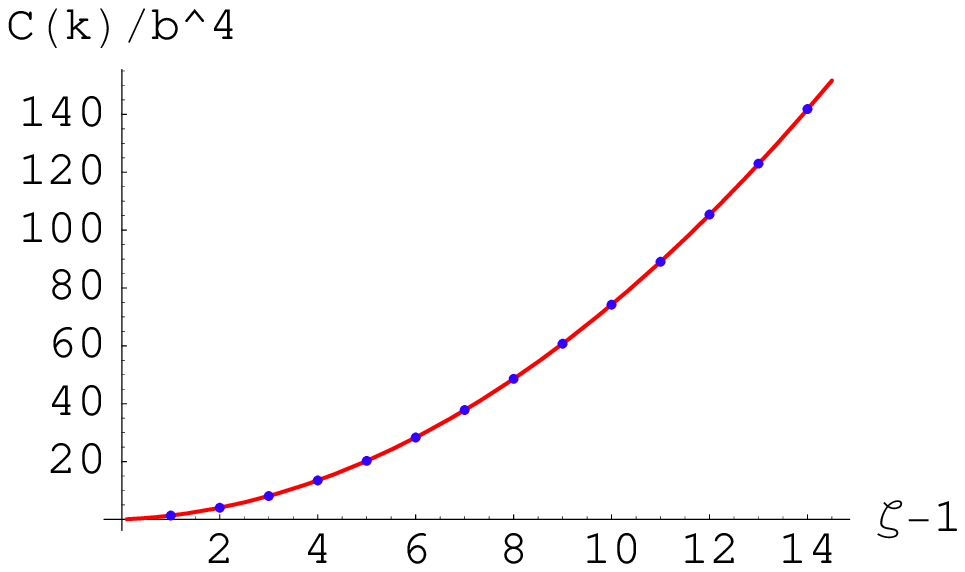}
\caption{Numerical comparison of the analytical result for the freely-jointed chain (solid line) with MC sampling of the discrete flexible chain as a function of interaction strength $K$; \textit{Left: }covariance $C_{k_1k_2k_3k_4}(K)$ for $(k_1,k_2,k_3,k_4)=(1,3,2,4)$; points: results from MC sampling of a chain of length $N=5$ with $10^8$ steps; line: analytical results using (\ref{fullcov}),(\ref{fisher}),(\ref{r12}),(\ref{m1}-\ref{m3}); \textit{Right: }Covariance $C_{1k_2k_3(k_2+1)}(K)$ for $K$ with varying overlap $\zeta=k_2-k_3+1$ from MC sampling (points); the solid line shows the first order linear approximation $C_{1k_2k_3(k_2+1)}(K)/b^4\approx\frac{2}{3}\zeta(\zeta-1)+\frac{8}{9}K (\zeta-1)^2$}
\label{hsnum}
\end{figure}

\begin{figure}
\centering
\includegraphics[width=6.5in]{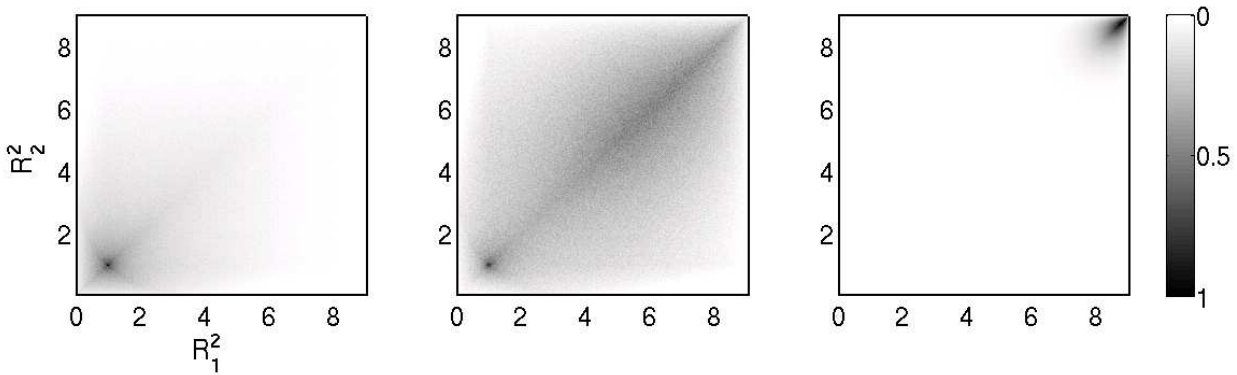}
\caption{Joint probability distribution function $P(R_{1}^2,R_{2}^2)$ for different coupling constants in the case $(k_1,k_2,k_3,k_4)=(1,3,2,4)$ obtained from MC sampling as in figure \ref{hsnum}. From left to right, the coupling constants are $K=0$, $K=K_{max}\approx 1.126$, $K=8$, gray and black colors indicate regions of higher probability (each panel normalized by $\max_{R1,R2} P(R_1^2,R_2^2)$).}
\label{dens}
\end{figure}

\begin{figure}
\centering
\includegraphics[width=3in]{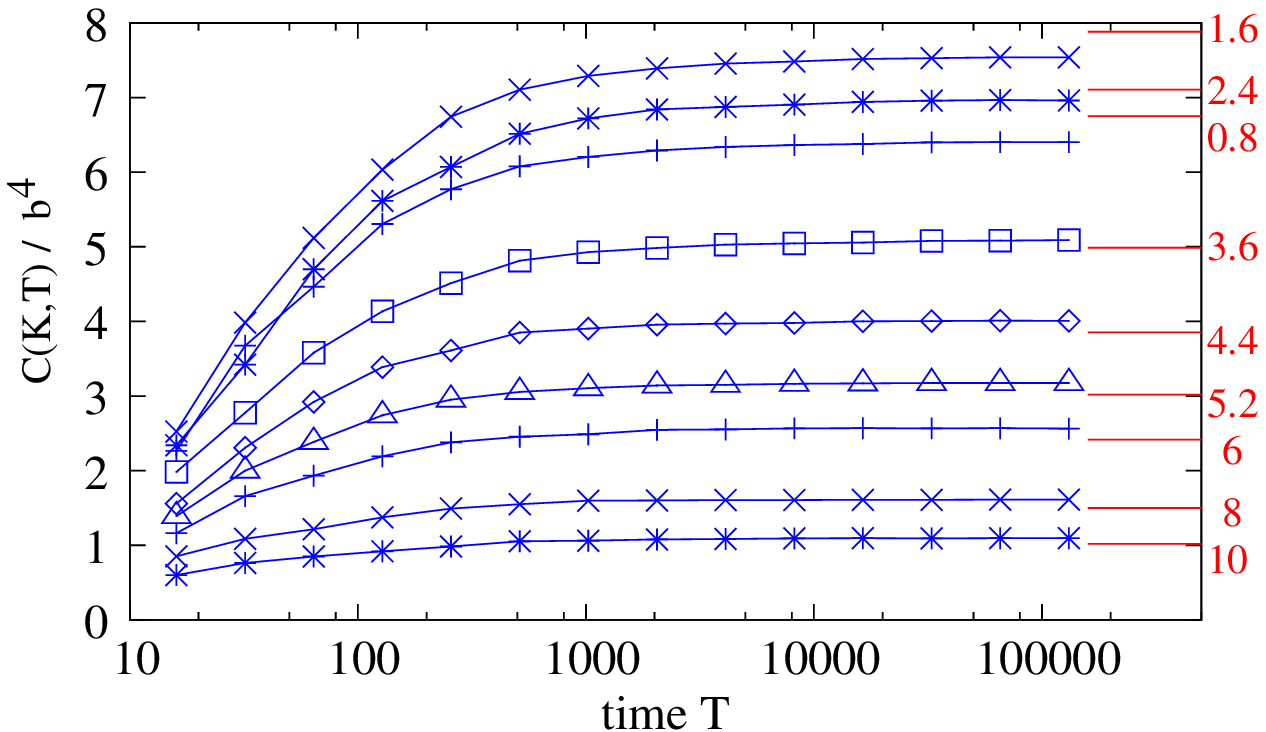}
\includegraphics[width=2.9in]{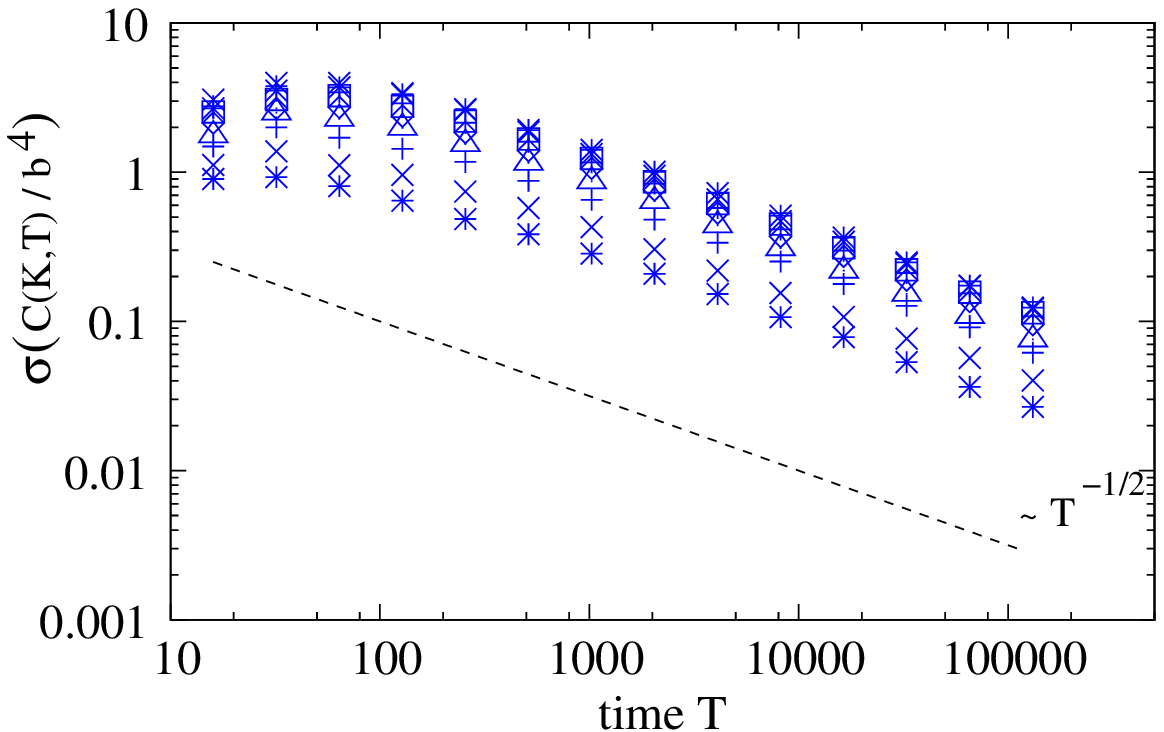}
\caption{\textit{Left:} Time convergence of running time average of the covariance $\mathcal{C}_{k_1k_2k_3k_4}(K,T)$ for different coupling constants $K$. Here, $(k_1,k_2,k_3,k_4)=(1,4,2,5)$ and $N=7$; the blue lines and points show the running time estimation of the covariance averaged over $1000$ independent initial conditions, the red lines indicate the exact result. \textit{Right: } Time convergence of the standard deviation of the distribution of covariances from different initial conditions. In both figures, the unit of time is $20dt$ (see text).}
\label{hsnum2}
\end{figure}

\begin{figure}
\centering
\includegraphics[width=4in]{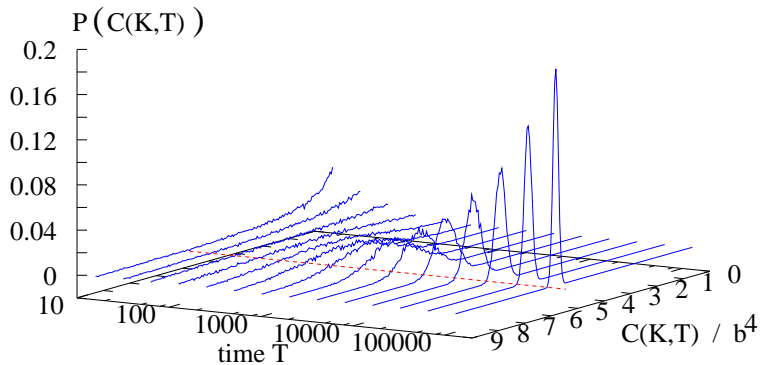}
\caption{Probability distribution function of the time averaged covariance function obtained from $10000$ independent initial conditions, $(k_1,k_2,k_3,k_4)=(1,5,4,6)$, $N=7$ and $K=0.4$. The red line indicated the analytically predicted value.  The unit of time is $20dt$ (see text).}
\label{hsnum3}
\end{figure}

\paragraph{ensemble averages}
We numerically evaluate the covariance using a Monte Carlo sampling of the classical one-dimensional Heisenberg model. A new trial configuration was obtained from the arbitrary reorientation of a randomly chosen spin. For small interaction strengths, such a sampling scheme achieves a high acceptance ratio. However, at stronger couplings, this ratio quickly drops. We evaluate the covariance $C(R_1^2,R_2^2)$ with $R_1=\sum_{i=k_1}^{k_2}\mathbf{r}_i$ and $R_1=\sum_{i=k_1}^{k_2}\mathbf{r}_i$, which we will denote from now on with the shorthand $C_{k_1k_2k_3k_4}(K)$ to indicate its dependence on the coupling strength $K$. On the left hand side of figure \ref{hsnum}, we show a comparison between the analytical and numerical results for Heisenberg model or, equivalently, the discrete flexible chain as a function of the coupling strength $K$ for a small chain with $N=5$ segments. We used $10^8$ Monte Carlo steps for each value of the coupling constant, and chose $(k_1,k_2,k_3,k_4)=(1,3,2,4)$, i.e. $\zeta=2$. Notice that for $K=0$, we recover the limit of the non-interacting freely-jointed chain, $C_{1324}(0)/b^4=4/3$. For small coupling $K$, the correction to the zero-coupling limit is linear, but the increase is sharper for larger $K$ and passes through a maximim. Then, for large values of $K$, the covariance decreases with $K$ towards zero. These results can be interpreted in view of the stiffening of the chain with increasing $K$: Whereas a large coupling constant initially tends to enhance cooperative motion among the segments, it also reduces the volume of the configurational space sampled on average as the energy penalty for overlapping monomers increases. This situation is schematically represented on the left hand side of figure \ref{hsnum}, where snapshots of typical configurations of a dynamical implementation of the discrete flexible chain (discussed in the next subsection) for values of $K$ below and above the maximum are shown. These can been understood more quantitatively by studying the joint probability distribution function $P(R_1^2,R_2^2)$ of the two distances. Figure \ref{dens} showns $P(R_1^2,R_2^2)$ in the previous case $(k_1,k_2,k_3,k_4)=(1,3,2,4)$ for different values of the coupling constant $K$: at zero interaction in the freely-jointed limit (left-hand panel), at $K=K_{max}\approx 1.126$ which maximizes  $C_{1324}(K)$ in figure \ref{hsnum} (middle panel), and at $K=8$ (right-hand panel). One observes that the transition $C_{1324}(K)\rightarrow 0$ for large $K$ goes along with a restriction to a relatively small part of configurational space associated to a stretched configuration of the polymer.\\{\ }\\
For small values of the coupling constant, the first order linear correction term can be estimated by performing simulations for varying overlap $\zeta=k_2-k_3+1$. On the right hand side of figure \ref{hsnum}, we evaluate the covariance function $C_{k_1k_2k_3k_4}(K)$ for $K=0.01$ fixed, and with $\zeta=2,...,15$ along with a corresponding increase of the chain size $L$. The results are fitted by the first order linear approximation function $C_{1k_2k_3(k_2+1)}(K)/b^4\approx\frac{2}{3}\zeta(\zeta-1)+\frac{8}{9}K (\zeta-1)^2$.
\paragraph{time averages}
Similarly to the non-interacting case, we would like to estimate the covariance from time averages using a dynamical implementation of the model. As outlined in section \ref{num1}, we implement a distance constrained chain of mass points, while adding an additional contribution to the potential energy function $V_{KP}=-K\sum_{i=1}^{N-1}\left( \hat{\mathbf{r}}_i\cdot \hat{\mathbf{r}}_{i+1}-1 \right)$ that accounts for the energy penalty for monomer overlap in the discrete flexible chain. In the case $K=0$, we recover the previous non-interacting case while for non-vanishing coupling values, the discrete flexible chain generates an orientation-dependent force. Here, we used a chain of $N=7$ segments to estimate the running-time average
\begin{eqnarray}
\mathcal{C}_{k_1k_2k_3k_4}(K,T)&=& \overline{{R}_1^2{R}_2^2}  -  \overline{{R}^2_1}\ \overline{{R}^2_2}
\end{eqnarray}
with $\overline{A(t)}=\frac{1}{T}\int_{t_0}^{t_0+T}dt'\ A(t')$ as defined previously. As an example, we chose $(k_1,k_2,k_3,k_4)=(1,4,2,5)$. The simulation parameters were the same as described in section \ref{num1} except for a smaller time-step $dt=0.02$ and an equilibration time of $5\cdot 10^3$ time units. Configurations were recorded for averaging every 20 time-steps. Figure \ref{hsnum2} shows the evolution of the running-time average on a logarithmic scale for different values of the coupling constant ($K=0.8,1.6,2.4,3.6,4.4,5.2,6,8,10$). Notice that these couplings include values below and above the coupling maximizing the covariance. For each $K$, the results were averaged over $1000$ initial conditions to avoid statistical dependence of the results on a particular initial condition for the trajectory. 
While all simulations converge towards values close to the analytical results indicated by the red lines on the left hand side of figure \ref{hsnum2}, one observes that within the time of simulation this value is not exactly reached, and the relative changes towards long times are small. This can be seen more quantitatively by taking a closer look at the standard deviation of the covariance at different points in time. On the right hand side of figure \ref{hsnum2}, we computed the standard deviation $\sigma$ of the covariance estimated from averages over different initial conditions. For long times, the variance decreases approximately as $T^{-1/2}$, indicating a slow change in time. For these timescales, increasing time is equivalent to increasing the number of statistically independent samples, such that the overall convergence behavior is Gaussian. This convergence behaviour is not limited to this particular observable, but can also obtained for other variables such as $\overline{R_1^2}$. An illustration of the convergence of this distribution is given in figure \ref{hsnum3}, where the probability distribution function of the covariance is shown as a function of time based on an estimation from $10000$ initial conditions. The scale of binning of the histograms at each time step is based on the large distribution of values at the smallest timescale. One observes that the onset of the scaling with $T^{-1/2}$ in figure \ref{hsnum2} is reflected here by the evolution toward distributions with a well defined peak and decreasing width at half maximum for long times.

\section{Summary and Outlook}
\label{outlook}
We introduced the covariance of distances as a simple measure allowing to characterize correlated and uncorrelated motion in polymers and biomolecules. In this work, we focussed on the exactly solvable cases only. We considered two highly simplified models of homogeneous polymers. These can be though of as some zeroth-order approximation to the real situation. A result which may seem surprising at first sight is that even a system without interacting forces can display
a non-vanishing covariance stemming merely from a geometric overlap in the distances. Moving to a more realistic system, the discrete flexible chain, we have shown that the results of the non-interacting chain can be recovered as a special case of the covariance for the system 
with interaction. Also, we have shown that for small values of the coupling constant, i.e. for small excluded-volume penalties or, equivalently, high temperatures, the correction to the results of the freely-jointed chain scales linear with the coupling constant. This effect is due to an increase of correlated motion. This trend is yet not monotonic. With increasing coupling constant, the covariance at small couplings reaches a maximum and then decreases to zero. As a consequence, there exists a finite value of $K$ for which the ensemble average of the covariance coincides with the value obtained at zero interaction. A simple interpretation of this observation is that though stronger coupling increases the cooperativity, it also reduces the configurational space sampled as the penalty of the excluded volume increases. Consequently, the average amplitude of distance fluctuation decreases with $K$.
\noindent In section \ref{kratky}, we also analyzed the convergence of the time average of the covariance measure towards the ensemble average obtained
from an exact calculation. While the quantitative difference between the predicted value and the results from numerical experiments certainly depend on the coupling to the stochastic heat bath, the qualitative results indicating that an algebraic decay in the variance would in principle occur for
any type of canonical sampling provided the time scale is long enough so as to decorrelate events along individual trajectories.
\noindent Regarding the situation in single molecule experiments, though our results have been in a highly idealized setting, some notable conclusions can be drawn. In single molecule FRET experiments, the acquisition time of the intensity signal is always limited either by instrumental factors or the destruction of the fluorescent marker by photobleaching. Accordingly, one can only expect to find meaningful results if the data acquisition time is much longer than the slowest timescale of the underlying dynamics of the molecule, and if the experiment can be sufficiently repeated
so as to allow one for averages over independent trajectories. In the present model where the interaction is dominated by a single parameter, 
it appears that the approach to convergence is not affected by the strength of interaction. Though the expectation value of the covariance varies 
with $K$, the approach towards this value based on running time averages is $K$-independent. The convergence behavior in this model is not sufficient
to draw conclusions on the physical state of the system (strongly/weakly interacting), but the mean value of the covariance is related (figure \ref{hsnum} in section \ref{kratky}). It therefore
appears compulsory for an experiment aiming to distinguish among different states to be calibrated on an absolute intensity scale.
\noindent From the results of the idealized models considered in this work, a promising future direction appears to be to apply the covariance analysis to more complex models of biopolymers where analytical approaches are not feasible anymore. Even if the ensemble average cannot be evaluated exactly, numerical experiments characterizing the dynamical convergence, similar to the ones performed on the polymer models in this article, are still feasible on long timescales. As an example, for proteins, a natural question arises whether the convergence of the covariance of distances within secondary structure elements differs significantly form the relaxation timescales on the level of tertiary structure. In particular, it would be promising to look at these cases under different physical conditions related to the folding process. The results of such a study might provide insight on how to choose, from a very large number of possibilities, particularly interesting locations for fluorescent labelling in single molecule spectroscopy, and hence getting a closer view on ''molecules at work'' beyond autocorrelation functions.

\subsection*{Acknowledgements}
J.-G.H. acknowledges support from the Coll\`ege Doctoral Franco-Japonais and the Region Rh\^one-Alpes. K.K.K. is supported by the ANR program GIMP ANR-05-BLAN-0029-01 and by the French Ministry of Research. It is a pleasure to acknowledge stimulating discussions with P. Borgnat,  H. Wendt (Lyon) and S. Takahashi (Osaka).

\appendix
\section{Derivation of 4-th order elements for the discrete flexible chain}
\label{appendix1}
In what follows, the derivation of the 4-th order elements required for the evaluation of the covariances in section \ref{kratky} is outlined in a more general way than required for the purpose of the present work by considering a chain with non-homogeneous couplings.

\subsection{The set of angles for integration}
\begin{figure}
\centering
\includegraphics[width=2.4in]{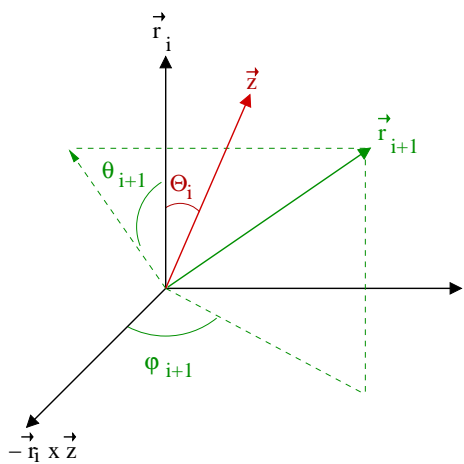}
\caption{illustration of the change of coordinates in (\ref{AppendixVecteurRj})}
\label{figrc}
\end{figure}

We choose the following coordinate system for integration (see illustration \ref{figrc}).  All unitary vectors $\r_{\ell}$ are marked by their polar angles with respect
to a fixed frame $\pa{ \Th_{\ell},\Phi_{\ell} }$. These are the so-called laboratory angles. To simplify the notation, we adopt the definition $\abs{\r_j}=1$ omitting the hat. We also choose a reference
vector $\r_i$ labelled by its laboratory angles. All other vectors $\r_j$ are marked by their polar coordinates  $\pa{\th_j,\varphi_j}$ with respect to the axial vector $\r_{j-1}$ if $j>i$ and  $\r_{j+1}$  if $j>i$. Then for $j>i$, the vector $\r_{j+1}$ is deduced from $\r_{j}$ by a rotation of $\r_j$ of axis $\v{z} \wedge \r_j $ and angle $\th_{j+1}$ followed by a rotation of axis $\r_j$ and angle $\varphi_{j+1}$. The situation is analogous for $j<i$.
A rotation of angle $\th$  and around a unitary axis $\v{a}$  acts on a vector $\v{x}$  according to
\beq
R_{\v{a},\th}\cdot \v{x}= \c{\th} \v{x}  + \pa{1-\c{\th}} \pa{\v{a}\cdot\v{x}} \v{a} + \s{\th} \pa{\v{a}\wedge \v{x}} \; .
\enq
An association of the both aforementioned rotations gives
\begin{eqnarray}
\r_{j+1}= \pa{\c{\Th_{j+1}} + \s{\th_{j+1}} \cot\pa{\Th_j} \c{\varphi_{j+1}}  } \r_j   - \s{\th_{j+1}} \pa{  \ba{c}    \s{\Phi_{j}} \s{\varphi_{j+1}}  \\
                                \c{\Phi_{j}} \s{\varphi_{j+1}} \\
                                \tf{ \c{\varphi_{j+1}} } { \s{\Th_j} }   \ea    } \; .
\label{AppendixVecteurRj}
\end{eqnarray}
(\ref{AppendixVecteurRj}) constitutes a recursion relation for the laboratory angles $\pa{\Th_j,\Phi_j}$ in terms of those located lower in the chain
and of the local angles $\pa{\th_j,\varphi_j}$. The longitudinal recurrence  is simple
\beq
\c{\Th_{j+1}}= \c{\th_{j+1}} \c{\Th_j} - \s{\th_{j+1}} \c{\varphi_{j+1}} \s{\Th_{j}} \; ,
\enq
whereas the transverse one is a little more involved:
\begin{eqnarray}
\r_{j+1}^{\perp}= \c{\th_{j+1}} \r_{j}^{\perp} + \s{\th_{j+1}}
\pa{\ba{c}
            \c{\varphi_{j+1}} \c{\Phi_j} \c{\Th_j}-\s{\varphi_{j+1}} \s{\Phi_{j}}  \\
            \c{\varphi_{j+1}} \s{\Phi_j} \c{\Th_j}+\s{\varphi_{j+1}} \c{\Phi_{j}}  \\   \ea }   \; .
\end{eqnarray}
These information are enough to compute the two and four-point correlation functions.
The latter will be computed for an inhomogeneous Hamiltonian
\beq
H=-k_BT \sul{j=1}{N-1} b_{j+1} {\r}_j\cdot {\r}_{j+1} \qquad \abs{{\r}_j}=1 \; .
\enq
The results of interest will follow either by taking the homogeneous limit, what we do in section \ref{kratky} where $b_i=K$, or by assuming that the couplings $\paa{b_i}$ are statistical
independent variables, and taking the ensemble average over such a distribution.  We start by re-deriving the known results for the two-point functions as the techniques involved there will be used for the higher-order correlators.

\subsection{The two-point function}

Let $\dd \om=\s{\th} \dd \th  \dd \varphi$  and $\dd \Om=\s{\Th} \dd \Th  \dd \Phi$. One has
\beq
\Int{}{} \dd \om  \ex{b \c{\th}}=
 4\pi \shc{b} \equiv  4\pi
\f{\sinh \pa {b} }{b}
 \; .
\enq
Then, for $i\leq j$, one has for the only non-trivial correlator
\begin{eqnarray}
\moy{r^z_i r^z_j} &=& \Int{}{}\f{\dd \om_1}{ 4\pi \shc{b_2} }
\dots \f{\dd \om_{i-1}}{ 4\pi \shc{b_i} }  \f{\dd \Om_i}{4\pi}
\f{\dd \om_{i+1}}{ 4\pi \shc{b_{i+1}} }
\dots \f{\dd \om_N }{ 4\pi \shc{b_N} }
  r^z_i r^z_j  \exp\paa{\sul{k=1}{i-1} b_{k+1} \c{\th_k} + \sul{k=i+1}{N} b_{k} \c{\th_k}  }   \nonumber   \\
\phantom{\moy{r^z_i r^z_j}}&=& \Int{}{} \f{\dd \Om_i}{4\pi}  \f{\dd \om_{i+1}}{ 4\pi \shc{b_{i+1}} }\dots \f{\dd \om_j}{ 4\pi \shc{b_{j}} }\ \ex{\sul{k=i+1}{j} b_{k} \c{\th_k}  }
  r^z_i r^z_j \; .
\end{eqnarray}
After this first trivial integration, in order to deal with the coupled integrals, we use the recursion of the angles. As $\Th_{j-1}$
is independent of $\pa{\th_j,\varphi_j}$ we can integrate on the latter variables.
We set
\beq
u \pa{b}=\Int{}{}  \f{\dd \om}{4\pi \shc{b} }  \c{\th} \ex{b \c{\th}}= \f{4\pi}{b^2}\f{b \cosh b - \sinh b }{\shc{b}}  \; ,
\enq
and since
$
\Int{}{}  \dd \om \s{\th} \c{\varphi} \ex{b \c{\th}}= 0  
$,
we get
\beq
\moy{r^z_i r^z_j}= u\pa{b_j}\moy{r^z_i r^z_{j-1}}=\ppl{k=i+1}{j}u\pa{b_k} \Int{}{} \f{\dd \Om}{4\pi} \cos^2\pa{\Th}  = \f{1}{3}\ppl{k=i+1}{j}u\pa{b_k}
\enq
recovering the result of Fisher\cite{fisher}. A similar calculation can be done for the four-point functions.
Here and in the following, we assume that $i\leq j \leq k \leq \ell$. Due to a trivial integration on the angles
located before the $i^{\eee{th}}$ site, one has
\beq
\moy{r^{a_i}_{i}r^{a_j}_{j}r^{a_k}_{k}r^{a_\ell}_{\ell}}_{b}=
\moy{  r^{ \wt{a_1} }_{1}r^{ \wt{a_j} }_{\wt{j}} r^{ \wt{a_k} }_{ \wt{k} } r^{ \wt{a_\ell} }_{ \wt{\ell} }    }_{\wt{b}} \ \ \ \ .
\enq
Where $\wt{j}=j-i+1$ and $\wt{b_k}=b_{i+k-1}$, $\wt{a_k}=a_{i+k-1}$. Hence, we can compute the correlators starting from the first site
and then, if needed, perform the aforementioned shifts.\\
We start with the XXZZ case. There, we choose $ \r_j$ to be the reference vector for all of the "moving frame" angles. Thus
\beq
\moy{r_1^x r_j^x r_k^z  r^z_{\ell} }= \Int{}{} \ppl{p=1}{j-1}\f{\dd \om_p \ex{b_{p+1} \c{\th_{p}}} }{4\pi \shc{b_{p+1}}  }   \f{\dd \Om_j }{4\pi}
\ppl{p=j+1}{\ell}\f{\dd \om_p \ex{b_{p} \c{\th_{p}}} }{ 4\pi \shc{b_{p}}  }  r_1^x r_j^x r_k^z  r^z_{\ell} \ \ \ \ .
\enq
The recursion relations for the angles read
\beqa
r_p^x  &=&  \c{\th_p} r_{p+1}^x + f_x\pa{\c{\varphi_p}, \s{\varphi_p}} \qquad  p \in \intn{1}{j-1}  \; , \\
r_p^z &=& \c{\th_p}   r_{p-1}^z + f_z\pa{\c{\varphi_p}, \s{\varphi_p}}   \qquad  p \in \intn{k+1}{\ell} \; .
\eeqa
The functions $f_{x,z}$ are linear in $\c{\varphi_p}$ and $\s{\varphi_p}$. Thence, their
 integrals over the azimuthal angle $\varphi_p$ give zero. We are thus led to the same recursion as for the two-point function. Eventually,
\beq
\moy{r_1^x r_j^x r_k^z  r^z_{\ell} }=  \ppl{p=2}{j} u\pa{b_p} \ppl{p=k+1}{\ell} u\pa{b_p}
\Int{}{}  \f{\dd \Om_j}{4\pi} \pa{r_j^x}^2  \pa{r_k^z}^2
\ppl{p=j+1}{k}\f{\dd \om_p \ex{b_{p} \c{\th_{p}}} }{ 4\pi \shc{b_{p}} }   \; .
\enq
The recurrence equation for the squared laboratory angles reads:
\begin{eqnarray}
\pa{r_k^z}^2=\pa{r_{k-1}^z}^2 \pa{  \cos^2 \th_k  -\sin^2 \th_k  \cos^2 \varphi_k   } 
+ \sin^2\th_k \cos^2 \varphi_k
-\f{1}{2}\sin 2\th_k \cos\varphi_k\sin 2 \Th_{k-1} \; .
\end{eqnarray}
The linear term in $\cos\varphi_k$ can be dropped after an integration over the azimuthal angle, so that after agreeing upon
\beqa
v\pa{b}= \Int{}{} \f{\dd \om}{4\pi \shc{b}}  \ex{b \c{\th}} \pa{\cos^2\th-  \sin^2\th \cos^2 \varphi}
=1- \f{3}{b^3} \paf{b \cosh b -\sinh b}{\shc{b}}  \; , \\
w\pa{b}= \pa{\Int{}{} \f{\dd \Om}{4\pi} \pa{r^x}^2 } \Int{}{} \f{\dd \om}{4\pi \shc{b}}  \ex{b \c{\th}} \sin^2\th \cos^2 \varphi
= \f{b \cosh b- \sinh b }{ 3 b^3 \shc{b} }\; \; ,
\eeqa
we get
\beq
I_k=v\pa{b_k} I_{k-1}+ w\pa{b_k}  \quad \eee{where} \quad  I_k =\Int{}{}  \f{\dd \Om_j}{4\pi} \pa{r_j^x}^2  \pa{r_k^z}^2
\ppl{p=j+1}{k}\f{\dd \om_p \ex{b_{p} \c{\th_{p}}} }{ 4\pi \shc{b_{p}} }    \; .
\enq
This is a linear recursion whose homogeneous solution is $I_k=\ppl{p=j+1}{k} v\pa{b_k} I_j$. Setting
$ \wt{I}_k= I_k\ppl{p=j+1}{k} v^{-1}\pa{b_p}$  we get
\beq
\wt{I}_k-\wt{I}_{k-1}=  \paf{w\pa{b_k}}{v\pa{b_k}} \ppl{p=j+1}{k-1} v^{-1}\pa{b_p} \; .
\enq
Thus, the overall solution reads
\beq
I_k= \f{1}{15}\ppl{p=j+1}{k} v\pa{b_k}  + \sul{p=j+1}{k} w\pa{b_p} \ppl{\ell=p+1}{k} v\pa{b_{\ell}} \; .
\enq
where we have written the explicit value of $I_j$. This expression can be slightly simplified if one takes averages
over the variables $b$. Assuming that these are independent random statistical variables, we get from averaging $[\cdot]$ over disorder
\begin{eqnarray}
I_k&=& \f{1}{15}\moyy{v\pa{b}}^{k-j}  + \moyy{w\pa{b}} \sul{p=j+1}{k} \moyy{v\pa{b}}^{k-p} = \f{1}{15} \moyy{v\pa{b}}^{k-j} + \moyy{w\pa{b}} \f{\moyy{v\pa{b}}^{k-j}-1}{\moyy{v\pa{b}}-1} \;.
\end{eqnarray}
Note that the results for a homogeneous Hamiltonian  follow from  the distribution $\de\pa{b_i-K}$. After performing the appropriate shifts we get
\beq
\moy{r_i^x r_j^x r_k^z r_{\ell}^z}= \moyy{u\pa{b}}^{j-i+\ell-k}
\paa{\f{1}{15} \moyy{v\pa{b}}^{k-j} + \moyy{w\pa{b}} \f{\moyy{v\pa{b}}^{k-j}-1}{\moyy{v\pa{b}}-1} } \; .
\enq
We now pass to the XZXZ and XZZX cases. The first reduction is identical with the previous case. Namely
\beq
\moy{r_1^x r_j^z r_k^x  r^z_{\ell} }=  \ppl{p=2}{j} u\pa{b_p} \ppl{p=k+1}{\ell} u\pa{b_p}
\Int{}{}  \f{\dd \Om_j}{4\pi} r_j^x r_j^z  r_k^z r_{\ell}^x
\ppl{p=j+1}{k}\f{\dd \om_p \ex{b_{p} \c{\th_{p}}} }{ 4\pi \shc{b_{p}} }   = \moy{r_1^x r_j^z r_k^z  r^x_{\ell} }\; .
\enq
These equalities are due to the fact that, for $a=x,z$, one has a relation of the type
\beq
r^{a}_\ell= \cos \th _{\ell} r^a_{\ell-1} + f^a\pa{\c{\varphi_{\ell}},\s{\varphi_{\ell}}} \quad
\mathrm{with} \quad \Int{0}{2\pi} d\varphi_{\ell}\ f^a\pa{\c{\varphi_{\ell}},\s{\varphi_{\ell}}}=0 \; .
\enq
Thus moving $r^z_{\ell}$ to the position $k$ produces the same coefficient as moving $r^x_{\ell}$ to the position $k$.
When writing a recursion for $r_k^z\, r_k^x$ it is enough to keep the terms that produce non-zero value after an integration
on the azimuthal variables $\varphi_{k}$. Thus, up to integration vanishing terms,
\beq
r_k^z r_k^x = \f{ r_{k-1}^z \, r_{k-1}^x }{ 2 } \pa{ 3 \cos^2{\th_k}  - 1} \; .
\enq
Setting
\beq
t\pa{b} = \f{1}{2} \Int{}{} \f{ \dd \om }{4\pi  \shc{b} } \ex{b \c{\th}}   \pa{ 3 \c{\th}  - 1} =
\f{1}{b^2} \pa{ b^2 +3 -3 \coth{b}} \; ,
\enq
we obtain
\beq
\moy{r_1^x r_j^z r_k^x r_{\ell}^z}= \f{1}{15} \ppl{p=2}{j} u\pa{b_p} \ppl{p=k+1}{\ell} u\pa{b_p}  \ppl{p=j+1}{k} t\pa{b_p} \;.
\enq
Thus performing the ensemble average as well as the shifts, we get
\beq
\moy{r_i^x r_j^z r_k^x r_{\ell}^z}= \f{1}{15} \moyy{u\pa{b}}^{j-i+\ell-k} \moyy{t\pa{b}}^{k-j} \; .
\enq


\section*{References}
\begin{thebibliography}{99}
\bibitem{ha}Roy R, Hohng S and Ha T 2008 {\it Nature Methods} \textbf{5} 507
\bibitem{karplus}Karplus M and McCammon J A 2002 {\it Nature Struc. Bio.} \textbf{9} 646
\bibitem{vangunsteren}van Gunsteren W F \textit{et al.} 2006 {\it Angew. Chem. Int. Ed. Engl.} \textbf{45} 4064
\bibitem{brown}Brown W M, Martin S, Pollock S N, Coutsias E A and Watson J-P 2008 {\it J. Chem. Phys.} \textbf{129} 064118
\bibitem{yang1}Yang H, Luo G, Karnchanaphanurach P, Louie T-M, Rech I, Cova S, Xun L and Xie X S 2003 {\it Science} \textbf{302} 262
\bibitem{yang2}Hanson J A, Duderstadt K, Watkins L P,  Bhattacharyya S, Brokaw J, Chu J-W and Yang H 2007 {\it Proc. Nat. Acad. Sci. USA} \textbf{104} 18055
\bibitem{hw}Hohng S, Joo C and Ha T 2004 {\it Biophys. J.} \textbf{87} 1328
\bibitem{watrob}Watrob H M, Pan C-P and Barkley M D 2003 {\it J. Am. Chem. Soc.} \textbf{125} 7336
\bibitem{doi}Doi M and Edwards S F 1988 \textit{The Theory of Polymers} (Oxford University Press)
\bibitem{bbk}Izaguirre J A, Catarello D P, Wozniak J P and Skeel R D 2001 {\it J. Chem. Phys.} \textbf{114} 2090
\bibitem{rattle}Allen M P and Tildesley D 1987 \textit{Computer Simulations of Liquids} (Oxford, Clarendon Press)
\bibitem{franklin}Franklin J and Doniach S 2006 {\it J. Chem. Phys.} \textbf{124} 154901
\bibitem{kp}Kratky O and Porod G 1949 \textit{Recl. Trav. Chim. Pays Bas} \textbf{68} 1106  
\bibitem{marko}Yan J, Kawamura R, and Marko J F 2005 \textit{Phys. Rev. E} \textbf{71} 061905 
\bibitem{mattis}Mattis D C 1987 {\it The Theory of Magnetism} vol. I, ch. 6.5f (New York, Springer)
\bibitem{takahashi}Nakamura H and Takahashi M 1994 {\it J. Phys. Soc. Jpn } \textbf{63} 2563
\bibitem{fisher}Fisher M E 1964 {\it Am. J. Phys. } \textbf{32} 343
\bibitem{tomita}Tomita H and Mashiyama H 1972 {\it Prog. Theor. Phys. } \textbf{48} 1133
\bibitem{mathematica}{Mathematica 5.0, Wolfram Inc.}

\end{thebibliography}
\end{document}